\documentclass[onecolumn,11.5pt]{article}
\usepackage[textwidth=6.5in,textheight=8.8in,headsep=80pt,footskip=72pt]{geometry}%

\usepackage{amsmath}
\usepackage{amssymb}
\usepackage[numbers,sort&compress]{natbib}

\usepackage{rotating}
\usepackage{url}
\usepackage{graphicx}
\usepackage{dcolumn}
\usepackage{bm}
\usepackage[usenames,dvipsnames]{xcolor} 
\usepackage{stmaryrd} 
\usepackage{bbm}
\usepackage[skip=5pt,font=small]{caption}

\newcommand{\new}[1]{{\color{black} {#1}}}

\begin{document} 

\title{Extracting the multi-timescale activity patterns of online financial markets}

\author{Teruyoshi Kobayashi,$^{1}$ $\;$ Anna Sapienza,$^{2}$ $\;$ Emilio Ferrara$^{2,*}$}

\date{\normalsize{$^1$\emph{Department of Economics, Kobe University, Kobe, Japan} \\
$^2$\emph{University of Southern California, Information Sciences Institute, Los Angeles (CA), USA}} \\
${}^*$Corresponding author: emiliofe@usc.edu\\
[2ex]
 \today}
\maketitle

\abstract{Online financial markets can be represented as complex systems where trading dynamics can be captured and characterized at different resolutions and time scales.
In this work, we develop a methodology based on non-negative tensor factorization (NTF) aimed at extracting and revealing the multi-timescale trading dynamics governing online financial systems. We demonstrate the advantage of our strategy first using synthetic data, and then on real-world data capturing all interbank transactions (over a million) occurred in an Italian online financial market (e-MID) between 2001 and 2015. 
Our results demonstrate how NTF can uncover hidden activity patterns that characterize groups of banks exhibiting different trading strategies (normal vs. early vs. flash trading, etc.). We further illustrate how our methodology can reveal ``crisis modalities'' in trading triggered by endogenous and exogenous system shocks: as an example, we reveal and characterize trading anomalies in the midst of the 2008 financial crisis. 
}

\section*{Introduction}

\noindent Social activity of individuals follows certain rhythms at different time scales. Many of the individual activities, such as sending e-mails and making phone calls, are likely to be done in particular time intervals within a day (i.e., diurnal or circadian cycles), and the total daily activity could heavily depend on the day of the week (i.e., weekly cycles)~\cite{Malmgren2008PNAS,Malmgren2009Science,Jo2011PlosOne,Jo2012NewJPhys,aledavood2015digital,aledavood2015daily}. In general, these social activity rhythms emerging at different time scales may be correlated with each other; for instance, it has been shown that face-to-face contacts between classmates in a primary school follow a common diurnal cycle driven by the daily class schedule~\cite{Stehle2011PLOS,gauvin2014detecting}, yet at the same time they would also share activity rhythms at longer scales such as weekly and monthly, reflecting the annual school schedule. 
 
  As social communications between humans form temporal social networks, financial transactions between banks also shape time-varying networks~\cite{May2008Nature,Haldane2011Nature,Barucca2016Soliton,Kobayashi2017arxiv,caccioli2018review}. In the interbank market, for instance, overnight bilateral lending and borrowing between banks organize temporal networks whose structure changes on a daily basis, because the overnight financial contracts last for only one day~\cite{Barucca2016Soliton,Kobayashi2017arxiv}. Thus, financial markets, similarly to social networks, can be interpreted as complex systems where each agent's activity should be captured and characterized at appropriate time scales~\cite{Barucca2016Soliton,Kobayashi2017arxiv}. In fact, many commonalities have recently been found between social communication patterns of humans and financial interaction patters of banks at particular temporal resolutions~\cite{Starnini2013PRL,Starnini2016SocNet,Kobayashi2017arxiv}.

 In recent years, \textit{non-negative tensor factorization} (NTF) has been frequently used to extract temporal activity patterns in various social contexts, such as face-to-face contacts~\cite{gauvin2014detecting,Sapienza2017arxiv}, Twitter posts~\cite{panisson2014mining} and players' matches in online games~\cite{sapienza2017NTFgame}. 
 Among these studies, Gauvin et al.~\cite{gauvin2014detecting} showed that NTF is highly effective in detecting diurnal rhythms of students' activity in a primary school~\cite{Cattuto2010PLOS,Stehle2011PLOS}. 
Characterizing diurnal rhythms lead the authors to uncover the multi-timescale community structure formed by classmates: students' temporal activity cycles, rather than the aggregated history of contacts, allows to reveal meaningful patterns that explain the complexity of children contact networks and communities.

Given the similarity between social and financial temporal dynamics, justified by underlying human factors, and the fact that multiple activity cycles are present at different frequencies in human social activities, our question is whether similar rhythms exist also in financial systems. 
In this work, we uncover hidden multi-timescale patterns of banks' activities in the Italian online interbank market, e-MID~\cite{emidHP}. In previous studies, it has been recognized that there exist activity patterns in banks' financial transactions at a particular time scale, such as inter-day~\cite{Kobayashi2017arxiv} and intra-day scales~\cite{Iori2008JEDC,Beaupain2008,kobayashi2017significant}. However, it is still unknown whether these patterns coexisting at different time scales are dependent on each other, in which case banks exhibiting a given interday activity pattern are likely to follow a particular intraday pattern.

 We employ NTF as a tool to detect multi-timescale patterns, in which banks' activities are captured by a tensor having three dimensions: the list of banks, time of the day, and date. 
 We will show that the PARAFAC decomposition of the 3-way tensor~\cite{Bro2003cc,kolda2009tensor} indicates that banks' trading activities can be classified into several patterns over the data period of 2001--2015. In particular, the NTF allows us to identify an anomalous pattern around the Lehman Brothers' collapse in September 2008. Our multi-timescale NTF approach reveals not only how banks facing the financial crisis changed their diurnal trading rhythms, but also on what dates such anomalies emerged. 
 
 \new{Over the last couple of years, many proposals for measuring and controlling systemic risk have been presented, yet most of them focus on the static nature of financial networks~\cite{Beale2011,Gai2011,kobayashi2014efficient}. However, as we show in this study of the Italian e-MID interbank market, various patterns and anomalies  systematically emerge at different time scales, which should be taken into account in the design of financial regulations.}
 \new{Importantly, even in a situation in which bank activity has certain patterns at intra- and inter-day scales, it remains of paramount importance to understand whether the intra-day patterns are independent of the inter-day patterns (i.e., mono-timescale patterns) or they are correlated with each other (i.e., multi-timescale patterns).  }
 Our framework to extract multi-timescale patterns advances the understanding of when and how banks react to  systemic shocks, and in the future it could contribute to controlling the spread of financial systemic risk~\cite{May2008Nature,Haldane2011Nature,GaiKapadia2010,Brummitt2015PRE,Battiston2016PNAS}.

\section*{Data}
\label{data}
We use the time-stamped data on bilateral interbank transactions occurred between January 2001 and December 2015 in the Italian online interbank market (e-MID). 
e-MID is an online platform  for financial institutions provided by the Italian company e-MID SIM S.p.A. based in Milan, Italy (henceforth, we call financial institutions as ``banks'' for brevity). Banks in need of lending or borrowing funds may find a counterpart by posting an order on the online platform, which makes e-MID a marketplace in which lenders and borrowers are matched. 
Each transaction record in this data represents when a loan contract agreement has been reached between two banks and how much funds have been lent (in million Euros). In addition, the data contains information about the type of each loan agreement, i.e., a lender-proposed transaction or a borrower-proposed transaction. We define a lender-proposed (resp., borrower-proposed) transaction to be a transaction proposed by a bank that lends to (resp., borrows from) its counterpart. 

As in the other interbank markets, the vast majority of transactions ($> 86$\%) in the e-MID market are overnight lending and borrowing (i.e., a loan contract lasts just for one day), while there are other types of transactions having longer maturity lengths such as two weeks, three month, etc.
To eliminate the influence of differences in maturity length, we focus on overnight transactions of unsecured Euro deposits labeled as ``ON'' (i.e., overnight) or ``ONL'' (i.e., overnight large, namely overnight transactions greater than 100 million Euros). 
Table~\ref{tab:summary_stats} summarizes the basic statistics of this dataset.

\begin{figure}[t]
     \centering
     \includegraphics[width=.99\columnwidth]{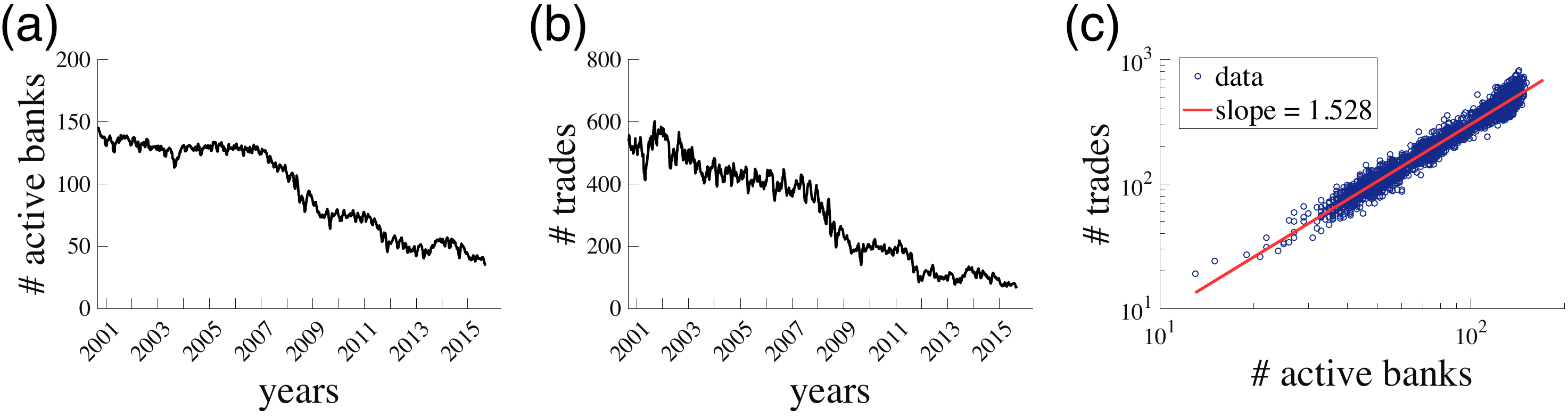}
     \caption{Daily activities in the e-MID market. 20-day moving average is shown \new{for (a) number of active banks having at least one transaction on a day and (b) number of trades. (c) Scaling relationship between the numbers of active banks and trades. Each dot represents a day.} }
     \label{fig:numbanks}
 \end{figure}

 There are 3,839 business days over the data period between Jan 2, 2001 and Dec 31, 2015, on each of which transactions are made between 8:00 and 18:00. There are 289 banks that conducted at least one transaction over this period: 194 of them are Italian banks while the rest are foreign banks. \new{Regardless of the nationality of banks, we examine the transactions conducted between 8:00--18:00 in Italian standard time. We do not consider a time difference between countries where the headquarters of the banks are located.} The total number of transactions (ON and ONL) that we use in the analysis amounts to 1,148,699. It should be noted that the 289 banks are not always active over the data period (\textit{cf.,} Fig.~\ref{fig:numbanks}a). In fact, the daily average of the number of participating banks is around 95, and the average number of daily transactions is approximately 300 (\textit{cf.,} Fig.~\ref{fig:numbanks}b). 
 \new{Although the time-series behaviors of the numbers of active banks and transactions are non-stationary, there is a stable scaling relationship between them (\textit{cf.,} Fig.~\ref{fig:numbanks}c)~\cite{Kobayashi2017arxiv}.} 
 The dataset is commercially available from e-MID SIM S.p.A (\url{http://www.e-mid.it/}). 
 
\begin{table}[]
            \caption{Summary statistics of the Italian interbank market, e-MID.}
    \centering        
    \begin{tabular}{lcccc}
    \hline
    \hline
   Data period & Jan 2, 2001 -- Dec 31, 2015 \\
   No.\ days & 3,839 \\
   Trading time & 8:00 -- 18:00 \\
   Maturity length & overnight (ON and ONL)\\
   No.\ banks (Italian banks) & 289 (194) \\
   No.\ total transactions  & 1,148,699 \\
   Average no.\ daily participants & 95.10 \\
   Average no.\ daily transactions & 299.22 \\
        \hline
    \end{tabular}\\
    \label{tab:summary_stats}
\end{table}

\section*{Method}


 As noted in the previous section, banks are not always active in the interbank market. At the daily scale, some banks have transactions only on a certain fraction of business days while having no transactions on the other days~\cite{Iori2008JEDC,Kobayashi2017arxiv}. If we look at intraday time scales, on the other hand, the frequency of a bank having at least one transaction is not uniformly distributed over the time intervals~\cite{Iori2008JEDC,Beaupain2008}. \new{These heterogeneities in temporal bank activity are illustrated in Fig.~\ref{fig:SI_hetero}.}    

Since the maturity of loan contracts is overnight, it is important to identify what are the days in which banks are more likely to be active. Conversely, since the participation of banks in the trading activity can change over the course of a day, it is also crucial to understand intraday activity changes. Due to the multi-scale nature of bank trading activity, directly studying the volume of transactions of each bank over time would not allow us to detect inter- and intra-day dynamics. Thus, we need to disentangle each bank's activity across the two dimensions (i.e., time scales): interday and intraday. To this aim, we represent our data in a three-dimensional array (i.e., tensor), whose entries represent the amount of trades each bank performs at a given time of a given day. Given this multi-dimensional representation, our main goal is to extract some meaningful correlated multi-timescale activity patterns related to groups of banks sharing a similar amount of transactions at the same time (respectively, intraday and interday). This can be achieved by means of \textit{non-negative tensor factorization} (NTF), as described in the following.
 

\subsection*{Non-negative tensor factorization (NTF) at different time scales}\label{sec:NTF_description}

\begin{figure}[t]
     \centering
     \includegraphics[width=.8\columnwidth]{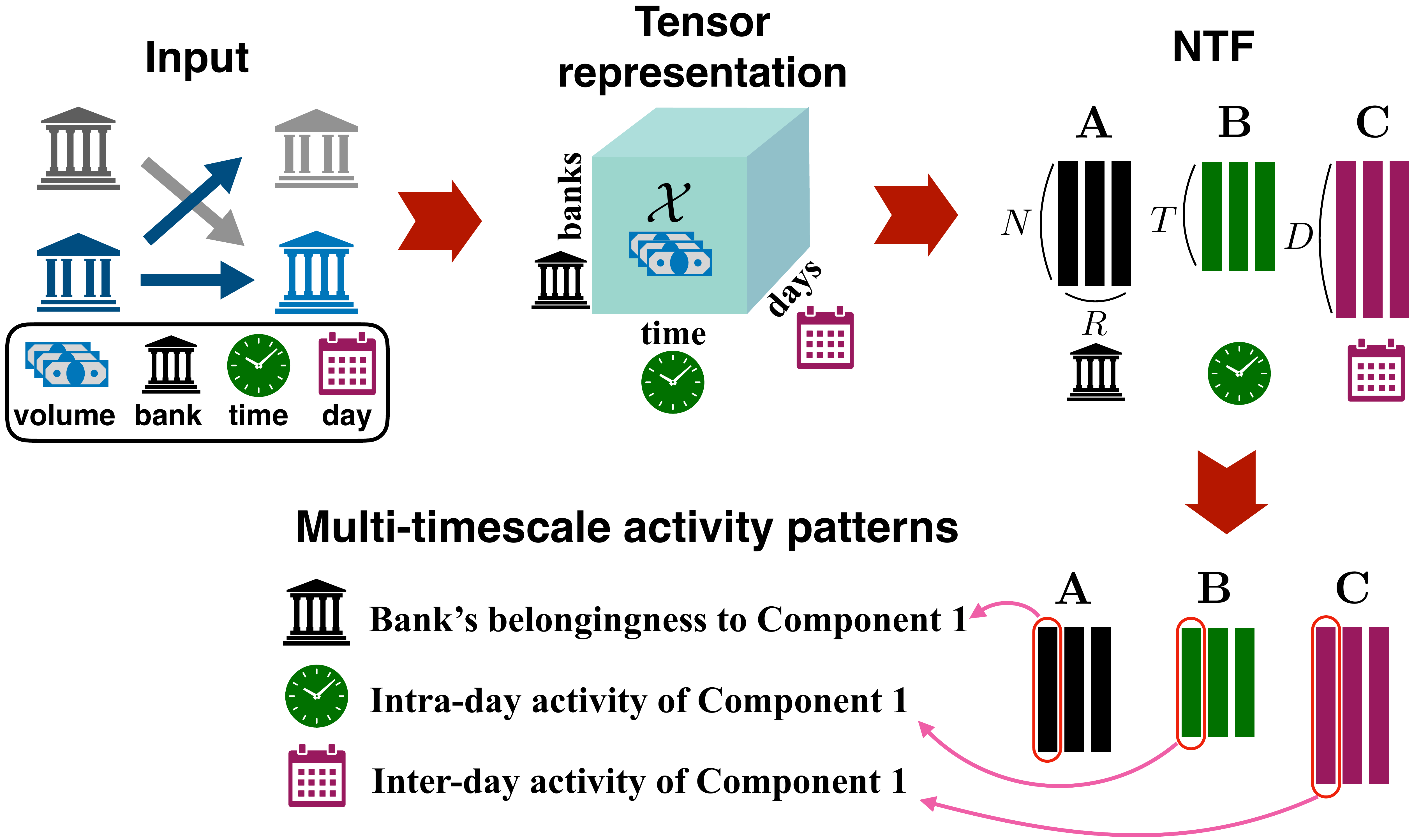}
     \caption{\new{Schematic of the implementation of Non-negative Tensor Factorization (NTF) framework.}}
     \label{fig:schematic}
 \end{figure}

 Let us consider a three-dimensional tensor $\mathcal{X}\in \mathbb{R}^{I\times J\times K}$, where $I=N$ is the number of banks, $J=T$ is the number of time intervals of bank activity in a day (here between 8:00 and 18:00), and $K=D$ is the number of days. The entry $x_{ijk}\in \mathbb{Z}_{+}$ of the tensor $\mathcal{X}$ denotes the total amounts of trades (in million Euros) conducted by the $i$-th bank during $j$-th time interval of the $k$-th day. 
 To extract the intra- and inter-day trading patterns characterizing banks with similar amount of transactions over time, we rely on the NTF, which decomposes the tensor $\mathcal{X}$ into the sum of several rank-one tensors, namely components \new{(\textit{cf.,} Fig.~\ref{fig:schematic}). } \new{For instance, if there exists a periodic inter-day pattern such as a two-day or a weekly cycle, then it would be captured by a component associated with the third dimension of the tensor.}
 
  Here, we implement the NTF by applying the  well-established method called PARAFAC (parallel factor analysis) or CANDECOMP (canonical decomposition) with non-negativity constraints~\cite{Bro2003cc,kolda2009tensor,lim2009nonnegative}. The decomposition can be written as
 \begin{align}
 \mathcal{X} \approx& \sum_{r=1}^{R}\mathbf{a}_{r}\circ \mathbf{b}_{r}\circ \mathbf{c}_{r},\label{eq:tensor} \\
 x_{ijk} \approx& \sum_{r=1}^{R}a_{ir}b_{jr}c_{kr},\label{eq:tensor_element}
 \end{align}   
where $R\in\mathbb{N}$ denotes the number of components (i.e., the rank of the tensor), and the operator $\circ$ represents outer product.  Eq.~\eqref{eq:tensor} is the so-called canonical polyadic decomposition (CP) of  a tensor $\mathcal{X}$, where $\mathbf{a}_{r}\in \mathbb{R}_{+}^{N}$, $\mathbf{b}_{r}\in \mathbb{R}_{+}^{T}$ and $\mathbf{c}_{r}\in \mathbb{R}_{+}^{D}$ represent $r$-th component factors that respectively encode the membership of a bank to the component, the intervals in a day and the days in which the component is active (i.e. $\mathbf{b}_{r}$ and $\mathbf{c}_{r}$ are the intra- and inter-day activity patterns of the groups of banks in $\mathbf{a}_{r}$).

 Let $\mathbf{A}\in \mathbb{R}_{+}^{N\times R}$, $\mathbf{B}\in \mathbb{R}_{+}^{T\times R}$ and $\mathbf{C}\in \mathbb{R}_{+}^{D\times R}$ denote the factor matrices, whose $r$-th columns are the vectors $\mathbf{a}_{r}$, $\mathbf{b}_{r}$ and $\mathbf{c}_{r}$, respectively.  The factor matrices 
$\mathbf{A}$, $\mathbf{B}$ and $\mathbf{C}$ are calculated by solving the following minimization problem with non-negativity constraints:
\begin{align}
 \min_{\mathbf{A}\geq 0,\mathbf{B}\geq 0, \mathbf{C}\geq 0} \lVert \mathcal{X} - \llbracket \mathbf{A},\mathbf{B},\mathbf{C}\rrbracket \rVert_{\rm F}^{2},
 \label{eq:parafac}
\end{align} 
 where $\llbracket \mathbf{A},\mathbf{B},\mathbf{C}\rrbracket$ represents the Kruscal form of the tensor decomposition (i.e., the RHS of Eq.~\eqref{eq:tensor}), and $\lVert \cdot\rVert_{\rm F}$ denotes the Frobenius norm.
 
 This minimization problem can be rewritten to perform the minimization with respect to one factor matrix at a time. Let $\mathbf{X}_{(n)}$ denote a matrix created by the mode-$n$ matricization (or flattening) of tensor $\mathcal{X}$, where each row of $\mathbf{X}_{(n)}$ consists of a vector corresponding to a given index of mode-$n$. That is, $\mathbf{X}_{(1)},\mathbf{X}_{(2)}$ and $\mathbf{X}_{(3)}$ are $N\times TD$, $T\times ND$ and $D\times NT$ matrices, respectively. They can be expressed in terms of factor matrices as follows:
  \begin{align}
      \mathbf{X}_{(1)} &= \mathbf{A}(\mathbf{C}\odot\mathbf{B})^\top, \\
      \mathbf{X_{(2)}} &= \mathbf{B}(\mathbf{C}\odot\mathbf{A})^\top, \\
      \mathbf{X_{(3)}} &= \mathbf{C}(\mathbf{B}\odot\mathbf{A})^\top, 
  \end{align}
  where $\odot$ denotes the Khatri--Rao product, which is the ``matching columnwise'' Kronecker product defined as $\mathbf{A}\odot\mathbf{B} = [\mathbf{a}_1\otimes\mathbf{b}_1,\ldots , \mathbf{a}_R\otimes\mathbf{b}_R]$. The minimization problem for PARAFAC decomposition Eq.~\eqref{eq:parafac} can be now reformulated as
  \begin{align}
      \left\{  \min_{\mathbf{A}\geq 0} \lVert \mathbf{X}_{(1)} - \mathbf{A}(\mathbf{C}\odot\mathbf{B})^\top \rVert_{\rm F}^{2}, \quad
        \min_{\mathbf{B}\geq 0} \lVert \mathbf{X}_{(2)} - \mathbf{B}(\mathbf{C}\odot\mathbf{A})^\top \rVert_{\rm F}^{2}, \quad
          \min_{\mathbf{C}\geq 0} \lVert \mathbf{X}_{(3)} - \mathbf{C}(\mathbf{B}\odot\mathbf{A})^\top \rVert_{\rm F}^{2},
      \right\}.
  \end{align}
  The minimizers of $\mathbf{A}$, $\mathbf{B}$ and $\mathbf{C}$ of this problem are computed using the non-negative alternate least squares method (ANLS) combined with the Block Principal Pivoting method (BPP) developed by \cite{kim2012fast}. Our implementation is based on the Tensor Toolbox~\cite{TTB_Software,TTB_Sparse} and the MATLAB codes  available from \cite{JinguKimHP}.

 \subsubsection*{Rank size}

To determine the number of components $R$ (i.e., rank size) used in the NTF model, we rely on the Core Consistency Diagnostic~\cite{Bro2003cc}.
  Given the tensor element $x_{ijk}$, we can rewrite Eq.~\eqref{eq:tensor_element} as
 \begin{align}
   x_{ijk} = \sum_{n=1}^{R}\sum_{m=1}^{R}\sum_{p=1}^{R}\lambda_{nmp}a_{in}b_{jm}c_{kp}, 
 \end{align}
 where $\lambda_{nmp}$ denotes the $(n,m,p)$ element of the superdiagonal binary tensor $\mathcal{L}$ (i.e., $\lambda_{nmp} = \delta_{nm}\delta_{mp}\delta_{np}$). Now let $g_{ijk}$ denote the $(i,j,k)$ element of the core tensor $\mathcal{G}$, which is obtained by fitting the data to the Tucker3 model~\cite{Bro2003cc}. In the Tucker3 model, the minimization problem reads
 \begin{align}
     \min_{\mathbf{G}} \lVert \mathbf{X} -\mathbf{A}\mathbf{G}(\mathbf{C}\otimes \mathbf{B})^{\top} \rVert_{\rm F}^{2},
 \end{align}
 where $\mathbf{X}$ is a $N\times TD$ matrix converted from tensor $\mathcal{X}$, and $\mathbf{G}$ is also a matricized version ($R_n\times R_{m}R_{p}$) of the core tensor $\mathcal{G}$. With the Tucker3 decomposition, the tensor is written as
 \begin{align}
     x_{ijk} = \sum_{n}^{R_n}\sum_{m}^{R_m}\sum_{p}^{R_p}g_{nmp}a_{in}b_{jm}c_{kp}. 
 \end{align}
 From the lemma of Bro and Kiers~\cite{Bro2003cc}, the core tensor $\mathcal{G}$ will be identical to the superdiagonal tensor $\mathcal{L}$ if the Tucker3 model is perfectly fitted and the factors have full column rank. If $\mathcal{G}$ is significantly different from the superdiagonal tensor $\mathcal{L}$, by contrast, it means that there are non-negligible interactions between factors and the PARAFAC model is not appropriate. Thus, we could assess whether the data should be fitted to the PARAFAC model or the Tucker3 model by measuring the distance between $\mathcal{G}$ and $\mathcal{L}$.
 
 We employ the Core Consistency (CC) value proposed by Bro and Kier~\cite{Bro2003cc} as a measure of the distance  between $\mathcal{G}$ and $\mathcal{L}$:
 \begin{align}
     {\rm CC} = 100\times \left( 1-\frac{\sum_{n=1}^{R}\sum_{m=1}^{R}\sum_{p=1}^{R}(g_{nmp}-\lambda_{nmp})^2}{R} \right),
 \end{align}
 where we imposed the constraint $R_n = R_m=R_p = R$ in implementing the Tucker3 decomposition.
$\rm CC$ takes $100$ if the PARAFAC model perfectly fits the data and less than $100$ (possibly a negative value) if the model does not fit perfectly. Note that a rise in $R$ will reduce fitting errors while increasing the possibility of over-fitting. In general, as $R$ increases, there arises more interactions between components, which makes the core tensor $\mathcal{G}$ far from superdiagonal, resulting in a low value of $\rm CC$. Therefore, it is reasonable to stop increasing $R$ before the symptom of over-fitting emerges.
 Our determination of $R$ is given as      
 \begin{align}
     R = \max R^\prime \in\{ R^\prime :\: {\rm CC}(R^{\prime})>L_{\rm cc},\: R^\prime\in \mathbb{N},  \},
     \label{eq:R}
 \end{align}
where ${\rm CC}(\cdot)$ denotes the core consistency as a function of the number of components, and $L_{\rm cc}$ is a threshold parameter. In line with \cite{Sapienza2017arxiv}, we set $L_{\rm cc} = 85$, and to minimize randomness created by a PARAFAC decomposition, we implement PARAFAC decomposition 20 times for a given rank size and use their mean as the value for ${\rm CC}(R^{\prime})$. Our implementation uses the MATLAB code available form \cite{corcondiaHP} which is based on the alternating nonnegativity-constrained least squares with block principal pivoting~\cite{papalexakis2015fastcorconia}.

\new{It should be noted that if one implements a low-dimensional clustering method, such as non-negative matrix factorization, separately at different time scales, then in general one would obtain different numbers of mono-timescale patterns at different scales. In that case, however, it would not be possible to know whether or not there exist multi-timescale patterns such that banks exhibiting a common intraday pattern also show a common interday pattern.}

 \section*{Results}

 \subsection*{Synthetic temporal financial markets}
 
\begin{figure}[t]
     \centering
     \includegraphics[width=.85\columnwidth]{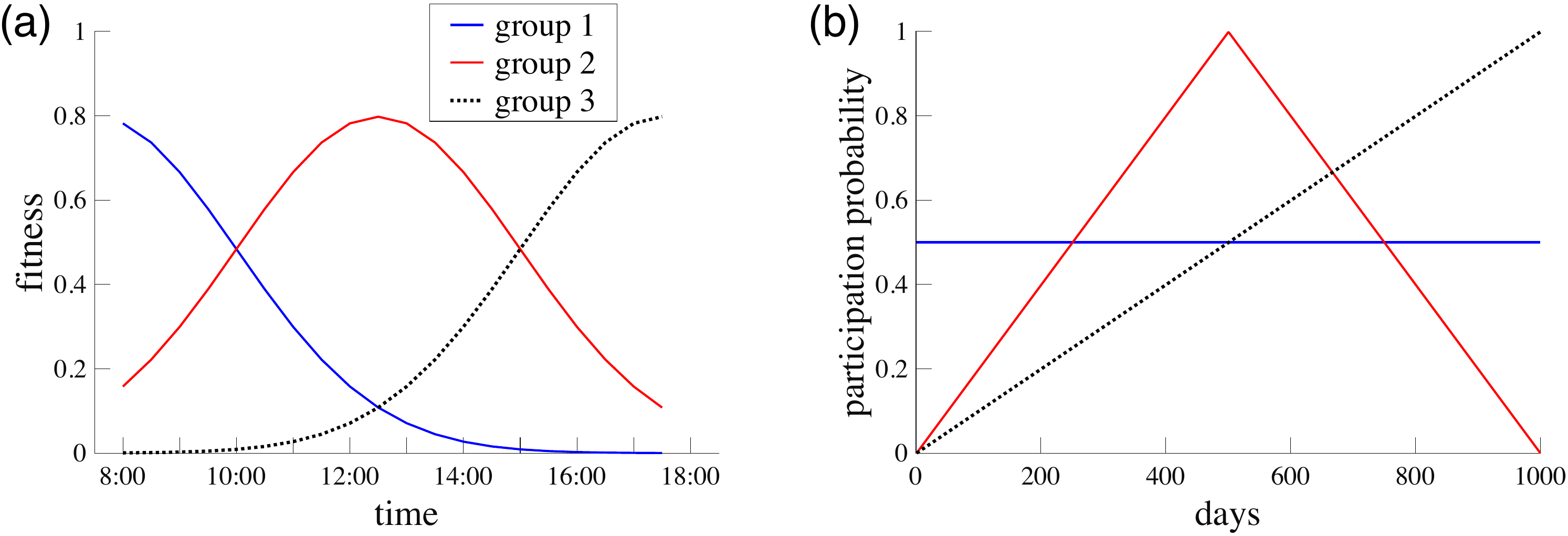}
     \caption{Predefined intra- and interday trading patterns for the synthetic financial market. (a) Predefined intraday activity patterns, represented by the fluctuation of fitness. (b) Predefined interday activity patterns, given by the participation probability.}
     \label{fig:pattern_synthetic}
 \end{figure}

  To validate the accuracy of our multi-timescale decomposition, we first implement NTF on a synthetic financial market whose properties are fully known \emph{ex ante}. Suppose that there are three groups of banks, each of which having particular intra- and inter-day trading patterns. First, we introduce \emph{intraday} trading patterns by considering the ``fitness'' of banks. In the fitness model, the probability that banks $i$ and $j$ trade at time $t$ is given by 
  \begin{align}
  p_{ij,t}=a_{i,t}a_{j,t},
  \end{align}
  where $a_{i,t}\in[0,1]$ denotes the fitness of bank $i$ within the time interval $[t-\Delta, t]$~\cite{DeMasi2006PRE,Kobayashi2017arxiv}. $\Delta$ is the length of a time interval (in a day) expressed in minutes. Differences in the intraday trading patterns across groups are then represented by different values of fitness. 
  Let us assume that the fitness evolves as illustrated in Fig.~\ref{fig:pattern_synthetic}a. Banks belonging to group 1 are most active in the early morning (blue solid), and banks in group 2 exhibit the highest trading activity around noon (red solid), and banks in group 3 are most active at the end of the day (black dotted). More specifically, the fitness value of a bank belonging to group $s$ at time interval $t$ is given as
  \begin{align}
    a_{i,t}^s = f(t;\mu_s,\sigma), \; t=1,\ldots,T,
  \end{align}
  where $f(t;\mu_s,\sigma)$ is the p.d.f. of normal distribution with mean $\mu_s$ and standard deviation $\sigma$. We set $(\mu_1,\mu_2,\mu_3) = (0,T/2,T)$ and $\sigma = T/4$.

   Second, the \emph{interday} trading patterns are captured by variations in a bank's participation probability, which is the probability that a bank participates in the market on a given day, apart from whether or not the bank is able to find a trading partner \textit{ex post}. For day $d=1,\ldots,D$, the participation probability of bank $i$, denoted by $q_{i,d}$, is given as (Fig.~\ref{fig:pattern_synthetic}b)
   \begin{align}
   q_{i,d}=
       \begin{cases}
        0.5 & \text{if $i\in$ group 1,}\\
        \frac{2}{D}(d-1) & \text{if $i\in$ group 2 and $d\leq D/2$,}\\ 
        \frac{-2}{D}(d-D) & \text{if $i\in$ group 2 and $d>D/2$,}\\  
        \frac{1}{D}(d-1) & \text{if $i\in$ group 3.}
       \end{cases}
   \end{align}
  For simplicity, we assume that trade volumes are identical for all trades so that the $(i,j,k)$ element of the 3-dimensional tensor is equal to the total number of trades that bank $i$ has during time interval $j$ on day $k$. \new{We will introduce a volume heterogeneity when analyzing the empirical data in the next section.} 
  We set $N=120$, $T = 20$ (i.e., resolution $\Delta=30$ for hours between 8:00 and 18:00 in each day) and $D=1000$.
  
   \begin{figure}[t]
     \centering
     \includegraphics[width=.85\columnwidth]{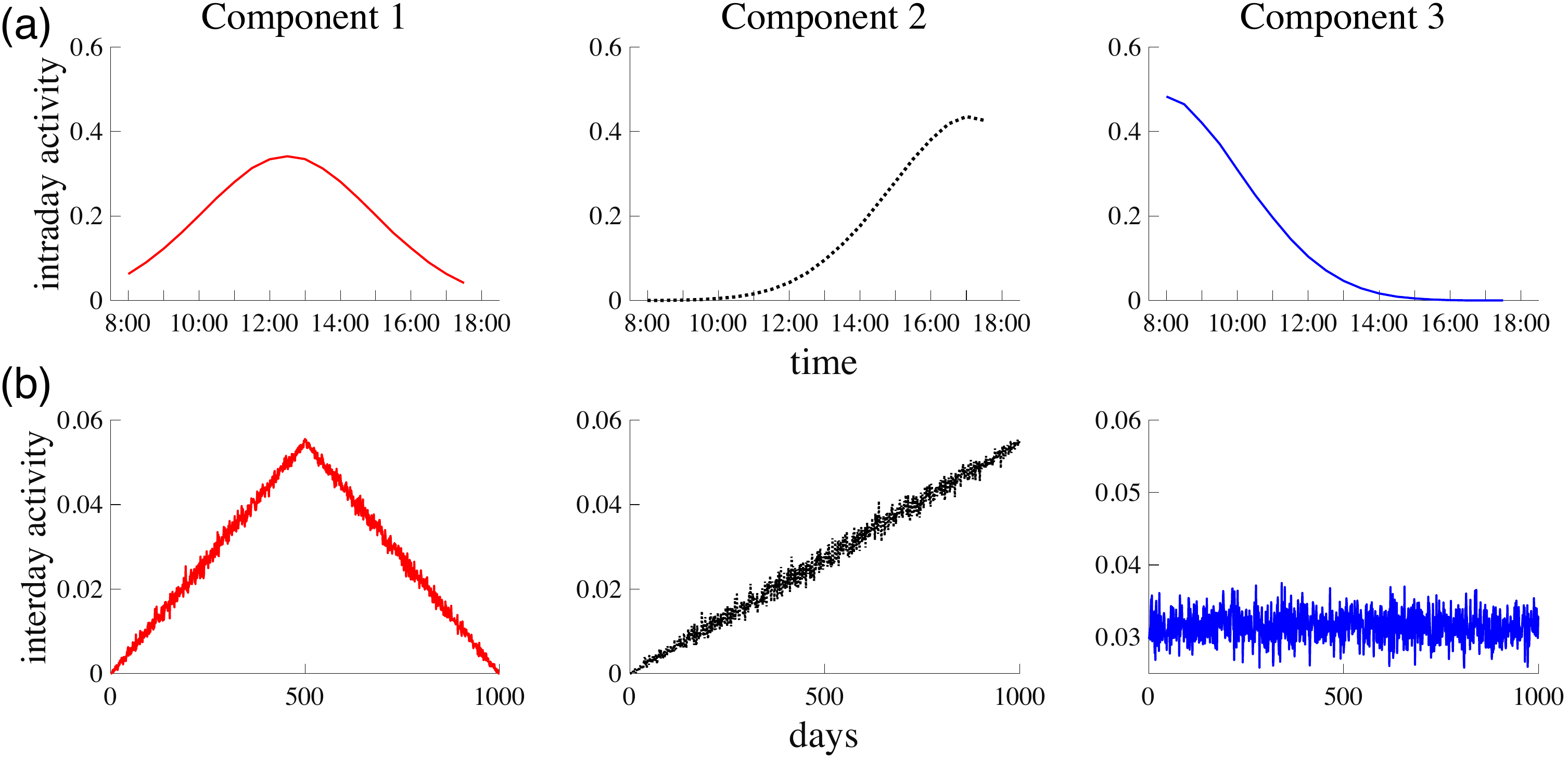}
     \caption{PARAFAC decomposition for the synthetic financial markets. (a) Intra-day activity for each component. Activity in time interval $j$ of component $r$ is given by $b_{jr}$. (b) Inter-day activity for each component. Activity of day $k$ of component $r$ is given by $c_{kr}$.}
     \label{fig:activity_synthetic}
 \end{figure}
   \begin{figure}[t]
     \centering
     \includegraphics[width=.35\columnwidth]{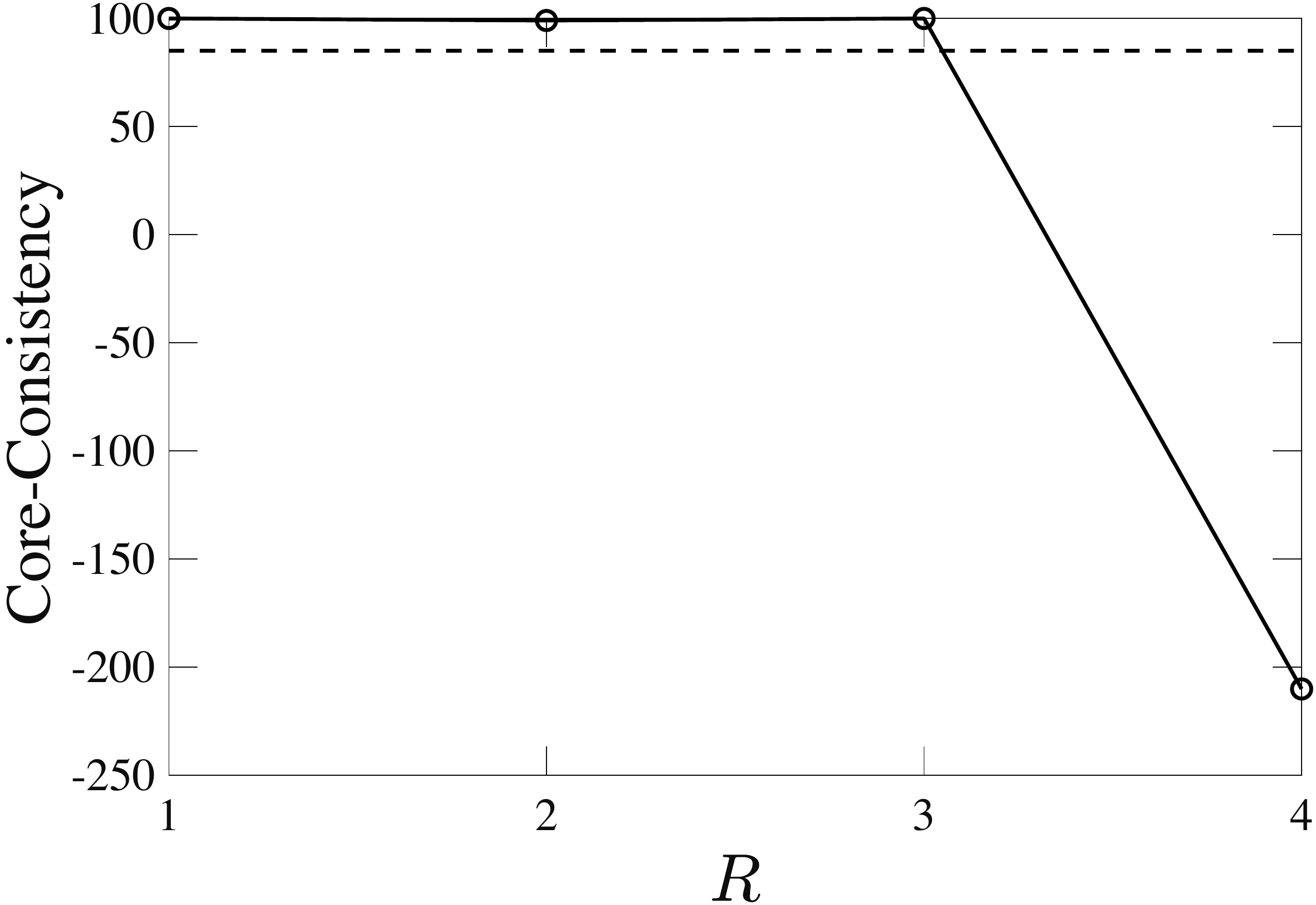}
     \caption{Core-Consistency for the synthetic markets. Dashed line denotes $L_{\rm cc} = 85$.}
     \label{fig:CC_synthetic}
 \end{figure}

 Using the above synthetic tensor, we can now implement our multi-timescale tensor factorization. The factorization of tensor $\mathcal{X}$ allows us to measure the activity of each individual element: a bank, an intra-day time interval, and a day. The activity of intra-day time interval $j$ of component $r$ is given by $b_{jr}$, and analogously the activity of day $k$ of component $r$ is represented by $c_{kr}$.
 Fig.~\ref{fig:activity_synthetic} illustrates that PARAFAC decomposition well extracts the true multi-timescale patterns. The core-consistency value strongly suggests $R=3$ since CC(4) takes a large negative value.

\subsection*{Empirical data} 
 
   Now we analyze empirical data. Given the accuracy of our method in the synthetic model, we expect the NTF method to extract latent multi-timescale patterns in the real-world financial market, e-MID. Here, we implement the $N\times T\times D$ tensor factorization described in Method, where we set intraday time resolution \new{(in minutes)} $\Delta =\{3,5,10,15,30,45,50\}$, and $D$ is set at the total number of business days between January 2001 and December 2015 (i.e., $D =$ 3,839). The number of active banks $N$, which traded at least once during the period, is 289. The Core-Consistency will reveal the number of components $R$, if any, which is equivalent to the the variety of existent activity patterns.

 \subsubsection*{Core-Consistency}
  
  \begin{figure}[t]
     \centering
     \includegraphics[width=.75\columnwidth]{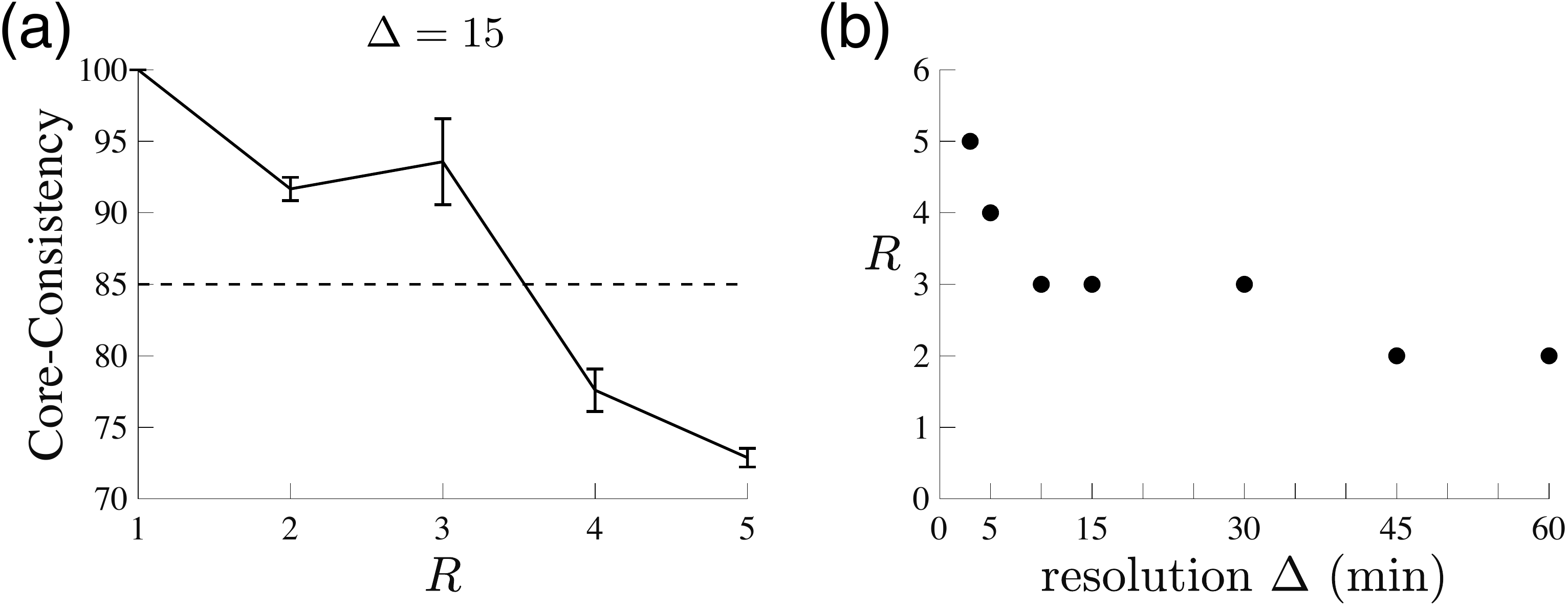}
     \caption{Determination of rank size. (a) Core-Consistency values for $\Delta = 15$. Error bar denotes the 95\% confidence interval of the mean Core-Consistency value, calculated by 20 runs of PARAFAC decomposition. (b) Selected value of $R$ for a given time resolution $\Delta \in \{3,5,10,15,30,45,60\}$.}
     \label{fig:CC}
 \end{figure} 
 
 The CC value for a given rank size is illustrated in Fig.~\ref{fig:CC}a for $\Delta =15$. The CC value becomes lower than 85 for $R\geq 4$, suggesting that $R=3$ should be selected. In general, imposing different intraday resolutions may lead us to choose different values of $R$ (Fig.~\ref{fig:CC}b). There is a tendency that a higher resolution requires a larger number of components for the PARAFAC decomposition to be appropriate. In the following, we employ $\Delta = 15$ and thereby $R=3$ as the benchmark case, the rationale of which will be discussed in the next subsection.

 \subsubsection*{Intra-day and inter-day activity}
  

   \begin{figure}[t]
     \centering
     \includegraphics[width=.9\columnwidth]{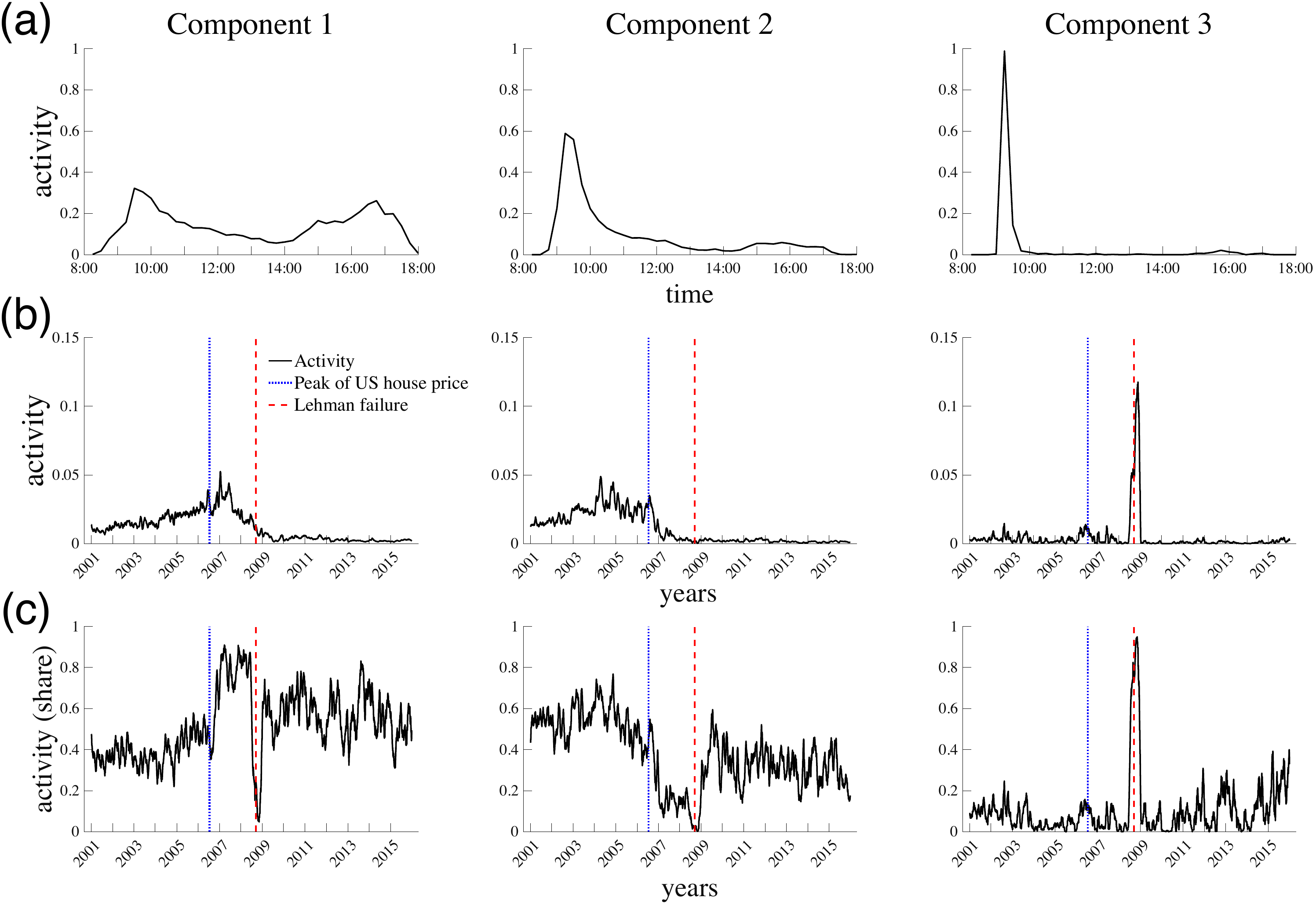}
     \caption{Intra- and inter-day activities. Peak of the US house price (July 2006) and Lehman collapse (September 2008) are indicated by blue dotted and red dashed, respectively. The data for the US house price is taken from FRED (S\&P/Case-Shiller U.S. National Home Price Index)~\cite{Case-Shiller_FRED} (a) Intra-day activity. (b) Inter-day activity. (c) Share of a component in the daily activity. In panels (b) and (c), 20-day moving average is shown. }
     \label{fig:activity_resol15}
 \end{figure}

  Fig.~\ref{fig:activity_resol15}a shows \emph{intra-} and \emph{inter-day} activities for $\Delta = 15$, in which case there are three different patterns. The intra-day activity of Component 1 is characterized by its bimodal pattern.\footnote{Since the assignment of component index is originally arbitrary in the NTF implementation, we re-assign index in the ascending order of cumulative intra-day activity between 8:00 and 10:00.} It has two peaks around 10:00 and 17:00. On the other hand, both Components 2 and 3 have a single distinct peak in the early morning, after which their activities are very low. However, the distribution of intraday activity of Component 3 is more skewed than that of Component 2, which makes Components 1 and 3 two polar cases and Component 2 in between them. We would call the trading patterns of Components 1, 2 and 3 as \emph{normal trading}, \emph{early trading}, and \emph{flash trading}, respectively.
  
  The activity patterns for different temporal resolutions are also presented in Fig.~\ref{fig:SI_activity}. For the case of a finer intra-day resolution (e.g., $\Delta=5$, Fig.~\ref{fig:SI_activity}a), there are more than three components, and the activity of an additional component (i.e., Component 4) turns out to be very similar to the flash-trading pattern we saw in Fig.~\ref{fig:activity_resol15}a. On the other hand, when the temporal resolution is low (e.g., $\Delta=45$, Fig.~\ref{fig:SI_activity}c), there are only two components and the flash-trading pattern is no longer detected. Given these observations, it is natural to set $\Delta=15$ as a benchmark case, where apparently independent intra-day patterns can be captured. It should be noted that the following results are not sensitive to the temporal resolution level as long as $R=3$ is to be selected (e.g., $\Delta=30$, Fig.~\ref{fig:SI_activity}b).

 Fig.~\ref{fig:activity_resol15}b illustrates the \emph{inter-day} activity of each component. For Components 1 and 2, the activity decreased radically during the global financial crisis of 2007--2009, which was initiated by a significant decline of the US house prices (the peak date is indicated by blue dotted). In contrast, the activity of Component 3 spiked around the collapse of Lehman Brothers in September 2008 \new{(indicated by red dashed)}. Fig.~\ref{fig:activity_resol15}c shows how the share of a given component evolved over time.
  
   These observations suggests that Component 3 captures the ``crisis mode'' of bank trading while Components 1 and 2 represent more normal trading patterns. The crisis-mode interpretation is reinforced by the intra-day trading pattern of Component 3, in which banks trade only in the very early morning. In fact, we will show that, although we did not use information about the directionality of trades, this anomaly comes from the fact that banks in need of liquidity tried to obtain loans, rather than to extend credit, in the midst of the financial crisis.

\subsubsection*{Banks' affiliation to components}
 
\begin{figure}[t]
     \centering
     \includegraphics[width=.98\columnwidth]{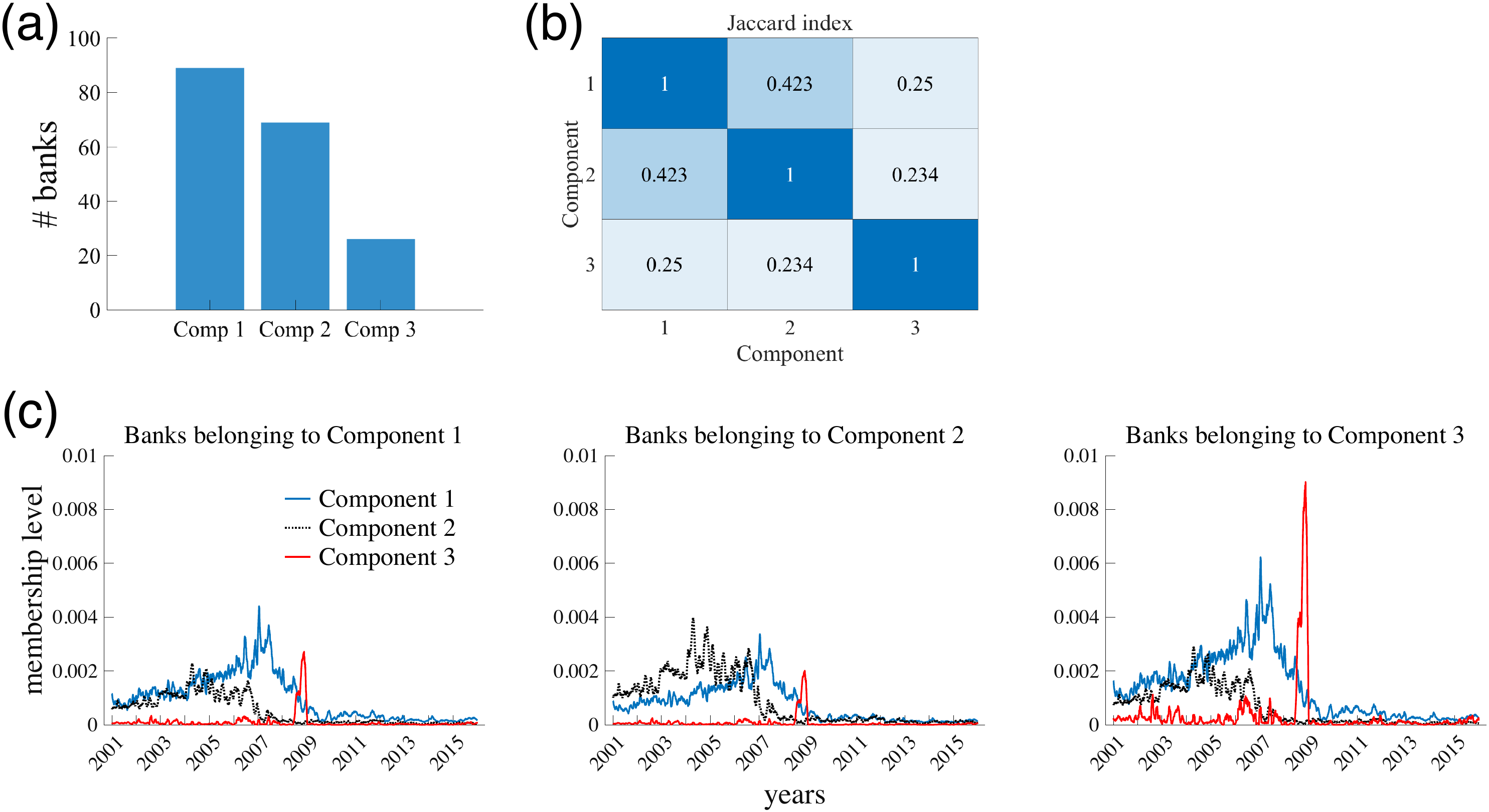}     
     \caption{Affiliation of banks to components. The set of banks belonging to component $r$ is given by $I_r = \{i^\prime\}$ for which $a_{i^\prime r}$ is within the 90th percentile of $\{a_{ir}\}_{i=1}^{N}$. (a) Number of banks that belong to a component. (b) Jaccard index for the overlap of banks belonging to multiple components. (c) Average membership level on a given day. Membership level for a bank is given by a row of $\mathbf{a}_{r}\mathbf{c}_{r}^{\top}$. }
     \label{fig:membership_90} 
 \end{figure} 
 
 Each element of vector $\mathbf{a}_r$ represents the extent to which a bank's trading pattern is captured by Component $r$. Thus, we can extract a subgroup of banks whose trading patterns are characterized by Component $r$ at least to some extent. 
 Here, the set of banks belonging to component $r$ is given by $I_r = \{i^\prime\}$ for which $a_{i^\prime r}$ is within the 90th percentile of $\{a_{ir}\}_{i=1}^{N}$ (Fig.~\ref{fig:membership_90}a). It should be noted that a bank may belong to multiple components if its trading pattern has multiple features (Fig.~\ref{fig:membership_90}b).
 
 The bank activity patterns emerging at the daily scale, captured by the row average of $\mathbf{a}_{r}\mathbf{c}_{r}^\top$, exhibit explicit differences across components (Fig.~\ref{fig:membership_90}c). While all types of banks have not been active since 2009, the normal-trading feature of banks belonging to Component 1 represents their behavior until 2008, and the trading pattern of banks belonging to Component 2 are dominated by the early-trading pattern up to 2006. On the other hand, the banks belonging to Component 3 exhibit daily patterns similar to the other types of banks prior to 2008, but suddenly their activity level spiked around the collapse of Lehman Brothers.
 
 \subsubsection*{Characterization of components}
 
  Aside from the intra- and inter-day trading patters, it would be of interest to see whether we may characterize each component by inherent attributes of the belonging banks. 
  In the e-MID data set, a trading bank is classified into either of the following four types: aggressor lender, quoter lender, aggressor borrower and quoter borrower. In the e-MID online platform, a bank posts a request for loans or a proposal for lending, and a loan contract is made if another bank accepts the posted request. A bank is called ``aggressor lender'' when the bank accepts a request for loan posted on the e-MID platform, and the bank posted the request and borrowed fund is called ``quoter borrower''. ``quoter lender'' and ``aggressor borrower'' can also be understood analogously. In addition, we can also ask to what extent the nationality of banks can explain the differences in trading patterns.

 \begin{figure}[t]
     \centering
     \includegraphics[width=.9\columnwidth]{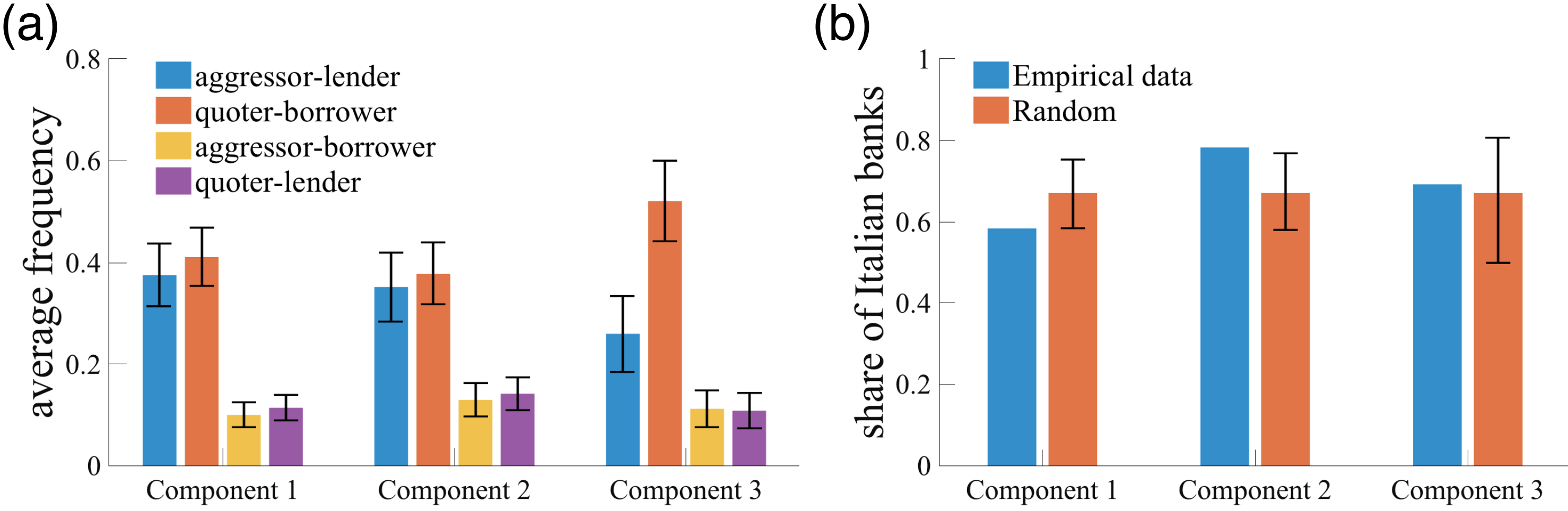}
     \caption{Role of bank attributes in the characterization of components. (a) Average frequency of a bank being an aggressor lender, a quoter borrower, an aggressor borrower and a quoter lender. Error bar denotes 95\% confidence interval. (b) Share of Italian banks in each group. ``Random" denotes the average share of Italian banks under random chance, and error bar denotes the 90\% interval calculated from a binomial distribution with parameter $(N_r,p)$, where $N_r$ is the number of banks belonging to component $r$ and $p\equiv \# \text{Italian banks}/N$.
     }
     \label{fig:type_component_90} 
 \end{figure}

 In fact, the majority of transactions in the e-MID market were done between aggressor lenders and quoter borrowers, which is a common property to all the component types (Fig.~\ref{fig:type_component_90}a). However, more than half of the banks belonging to Component 3 are quoter borrowers and the share of aggressor lenders is significantly lower than that. As for the nationality, the share of Italian banks among all the banks belonging to Component 1 is significantly low while the corresponding fraction is significantly large for the banks in Component 2 (Fig.~\ref{fig:type_component_90}b) at 90\% statistical significance, respectively. This suggests that foreign banks (i.e., banks from countries outside Italy) are likely to exhibit the normal-trading pattern, and Italian banks are more likely to employ the early-trading pattern than foreign banks do.

 These observations give us important implications about the multi-timescale trading patterns in the financial market.
 First, flash trading, which represents the intra-day trading pattern of Component 3, can be attributed mostly to quoter borrowers. Recall that flash trading was observed mostly in the period around the Lehman collapse in September 2008, but we do not know \textit{ex ante} whether such a trading pattern was driven either by demand or supply. The fact that Component 3 is largely attributed to quoter borrowers suggests that the flash-trading pattern during the financial crisis was conducted by banks that attempted to obtain liquidity as early as possible by posting quotes for loans. This may be regarded as evidence that some banks in fact faced a serious liquidity shortage at the time of the Lehman collapse~\cite{Brunnermeier2009JEP,Allen2010}. Our multi-timescale NTF approach reveals not only how banks reacted to the fear of liquidity shortage (i.e., intra-day pattern), but also specific dates on which the fear was most evident (i.e., inter-day pattern). 
 
 Second, differences in the nationality of banks can lead to the variety of trading patterns. It turns out that foreign banks (Italian banks) are more likely to employ normal trading (early trading) than early trading (normal trading). On the other hand, the source of the crisis modality cannot be explained by the nationality of banks (Fig.~\ref{fig:type_component_90}b), suggesting that banks had similar chance of being in a crisis mode regardless of their nationalities.

\section*{Discussion}
 In this work, we presented an analytic framework based on  \textit{non-negative tensor factorization} (NTF) to extract temporal activity patterns of financial systems. Despite  previous studies on online financial markets recognized the existence of trading activity patterns  at specific time scales (e.g., inter-day~\cite{Kobayashi2017arxiv}, or intra-day~\cite{Iori2008JEDC,Beaupain2008,kobayashi2017significant}), we demonstrated---to the best of our knowledge for the first time---that activity patterns coexist at different time scales and depend upon each other. Our  methodology allowed us to uncover the hidden multi-timescale patterns of trading activities in an online interbank market (e-MID~\cite{emidHP}). 

Within our framework,  banks' activities were represented by means of a tensor of three dimensions, namely the list of banks, time of the day, and date. Leveraging the power of NTF (based on the PARAFAC decomposition of the 3-way tensor~\cite{Bro2003cc,kolda2009tensor}), we uncovered the multi-timescale nature of financial trading dynamics in e-MID, suggesting that banks' trading patterns could be classified into subgroups exhibiting significantly different multi-timescale patterns. Our framework also allowed us to attribute roles to banking institutions (aggressor lender/borrower, or quoter lender/borrower) based on their trading patterns, yielding interpretable analytic insights.

By modeling the multi-timescale dynamics occurred over the period covered by our data (2001--2015), the proposed NTF-based framework allowed us to identify trading anomalies in the midst of the financial crisis. For example, around the time of Lehman Brothers' collapse in September 2008, our approach showed how banks changed their trading rhythms, and on what dates such anomalies emerged.

Understanding trading patterns in financial systems is of fundamental importance for many  reasons: first, a rigorous characterization of such systems' trading dynamics could help central banks to effectively intervene in the interbank market. In turn, this  would contribute to reducing systemic risk in the financial system as a whole. 
Financial linkages created by bilateral transactions between banks can lead to a global network of interconnected risk, in which a failure of one bank can immediately spread over the entire system~\cite{May2008Nature,GaiKapadia2010,Brummitt2015PRE}. 
Extracting multi-timescale patterns of financial markets allows for a better understanding of when and how banks react to  systemic shocks.

\new{Of course, the current approach has certain limitations. First, the tensor representation of interbank transactions does not capture information about the network aspect of the interbank market. While the current NTF takes bank ID, time and day as inputs, the interconnectivity between banks is still ignored. Given that many previous studies revealed that interbank networks contain rich information regarding systemic risk~\cite{May2008Nature,GaiKapadia2010,Brummitt2015PRE}, including structural information in the framework could improve the results.

Second, the current analysis did not fully explore the role of multi-scale activity patterns in propagating systemic risk. It would be interesting to examine to what extent each of the three activity patterns could promote or prevent financial contagion of bank defaults. To the best of our knowledge, the multi-timescale aspect of the source of financial systemic risk has not been investigated so far. This appears as a promising future direction of this line of research.

Finally, while the current work focused on the Italian e-MID market, this  is just a small part of the entire financial system. There are many other interbank markets in different countries, and there are also other types of interbank transactions such as trading of government bonds and credit default swaps. An important question to ask would therefore be: are the multi-timescale patterns found in our analysis ubiquitous or they are specific of markets that obey trading dynamics typical of the e-MID market?}

This work lays out the foundations for other researchers to study how  patterns and anomalies emerge in financial systems at different temporal resolutions:  in the future, such a framework could contribute to controlling the spread of financial systemic risk~\cite{May2008Nature,Haldane2011Nature,GaiKapadia2010,Brummitt2015PRE,Battiston2016PNAS}. 
\new{It would be interesting to} investigate whether our method could be used both in predictive and prescriptive ways to help determine what could happen under certain risk conditions, as well as to provide possible  intervention strategies prior to shocks or cascading collapses in financial systems.




\section*{Acknowledgements}
TK acknowledges financial support from the Japan Society for the Promotion of Science Grants no. 15H05729 and 16K03551. 
EF is grateful to DARPA for support (grant \#D16AP00115). This project does not necessarily reflect the position/policy of the Government; no official endorsement should be inferred. Approved for public release; unlimited distribution.

\section*{Author contributions}
TK and EF conceived the research and wrote the manuscript. TK performed the analysis. TK, AS and EF discussed the results and reviewed the manuscript. 

\section*{Competing interests} 
The authors declare no competing interests.

\clearpage


\clearpage

\setcounter{section}{0}
\setcounter{table}{0}
\setcounter{equation}{0}
\setcounter{figure}{0}
\setcounter{page}{1}
     
\renewcommand{\thetable}{S\arabic{table}}
\renewcommand{\thefigure}{S\arabic{figure}}
\renewcommand{\thesection}{S\arabic{section}}
\renewcommand{\theequation}{S\arabic{equation}}

\fontsize{16pt}{18pt}\selectfont

{\center{{\textbf{Supplementary Information}}  \\
\vspace{.5cm}
``Extracting the multi-timescale activity patterns of online financial markets'' \\
\vspace{.5cm}
\fontsize{13pt}{15pt}\selectfont
{Teruyoshi Kobayashi, Anna Sapienza and Emilio Ferrara}\\
}}

\vspace{1cm}
\fontsize{10pt}{12pt}\selectfont

\section{Temporal heterogeneity of bank activity}
\begin{figure}[thb]
     \centering
     \includegraphics[width=.95\columnwidth]{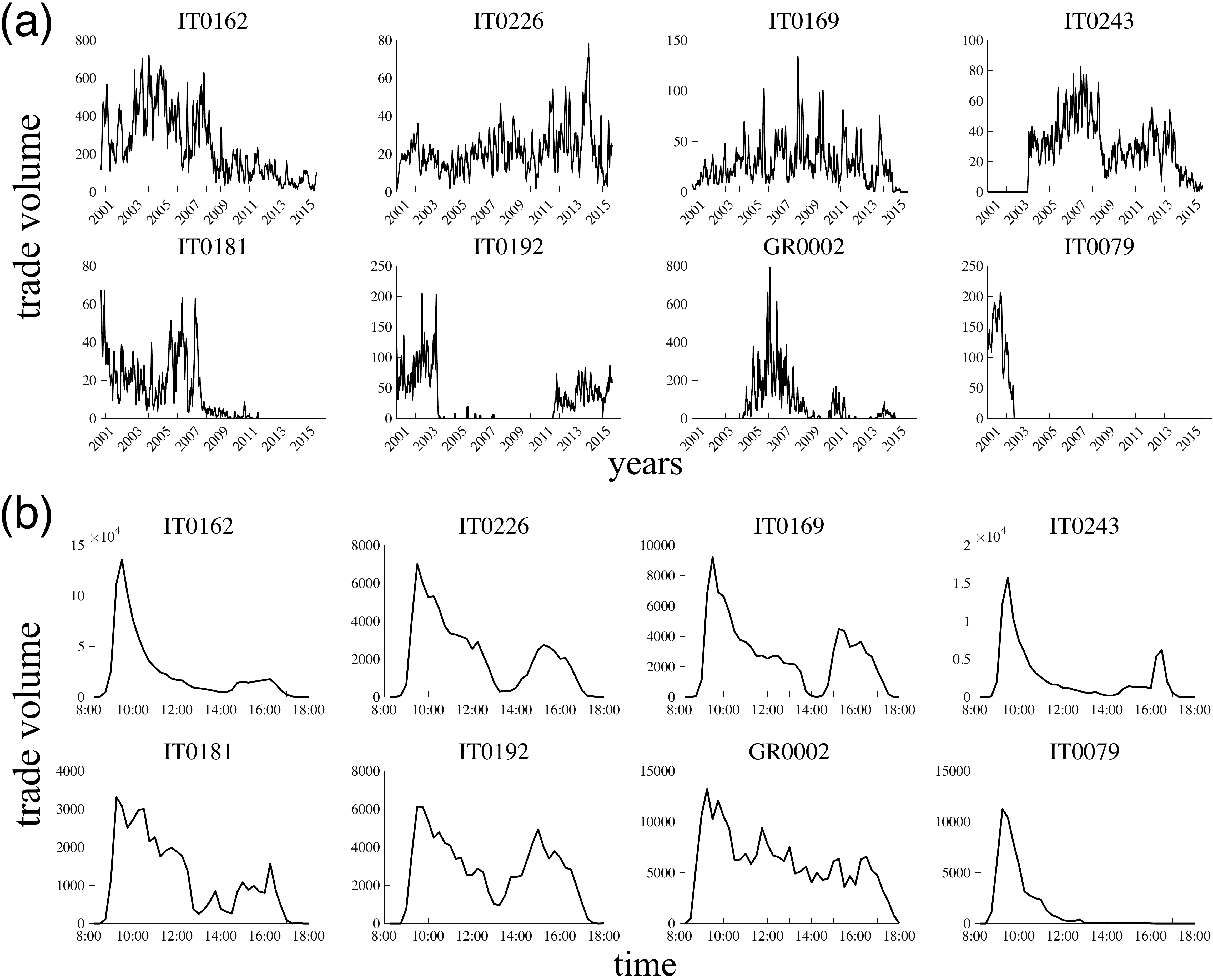}
     \caption{\footnotesize{\new{Bank activity at interday and intraday scales. Bank ID is annotated at the top of each panel. (a) 20-day moving average for the total volumes of transactions (in million Euros) conducted in each day. (b) Total amount of transactions conducted by a bank in each time interval (i.e., 15 min).}}}
     \label{fig:SI_hetero} 
 \end{figure}

\new{
 In Fig.~\ref{fig:SI_hetero}, we illustrate the heterogeneity of bank activity at inter- and intraday scales. We pick eight banks whose number of participated days are ranked 1st, 10th, 30th, 50th, 80th, 100th, 150th and 200th. The banks' activities are measured by their total volumes of transactions within a given day or a given time interval. }

\bigskip

\section{Intra- and inter-day activity of banks}
\fontsize{10pt}{12pt}\selectfont

\begin{figure}[ht]
     \centering
     \includegraphics[width=.95\columnwidth]{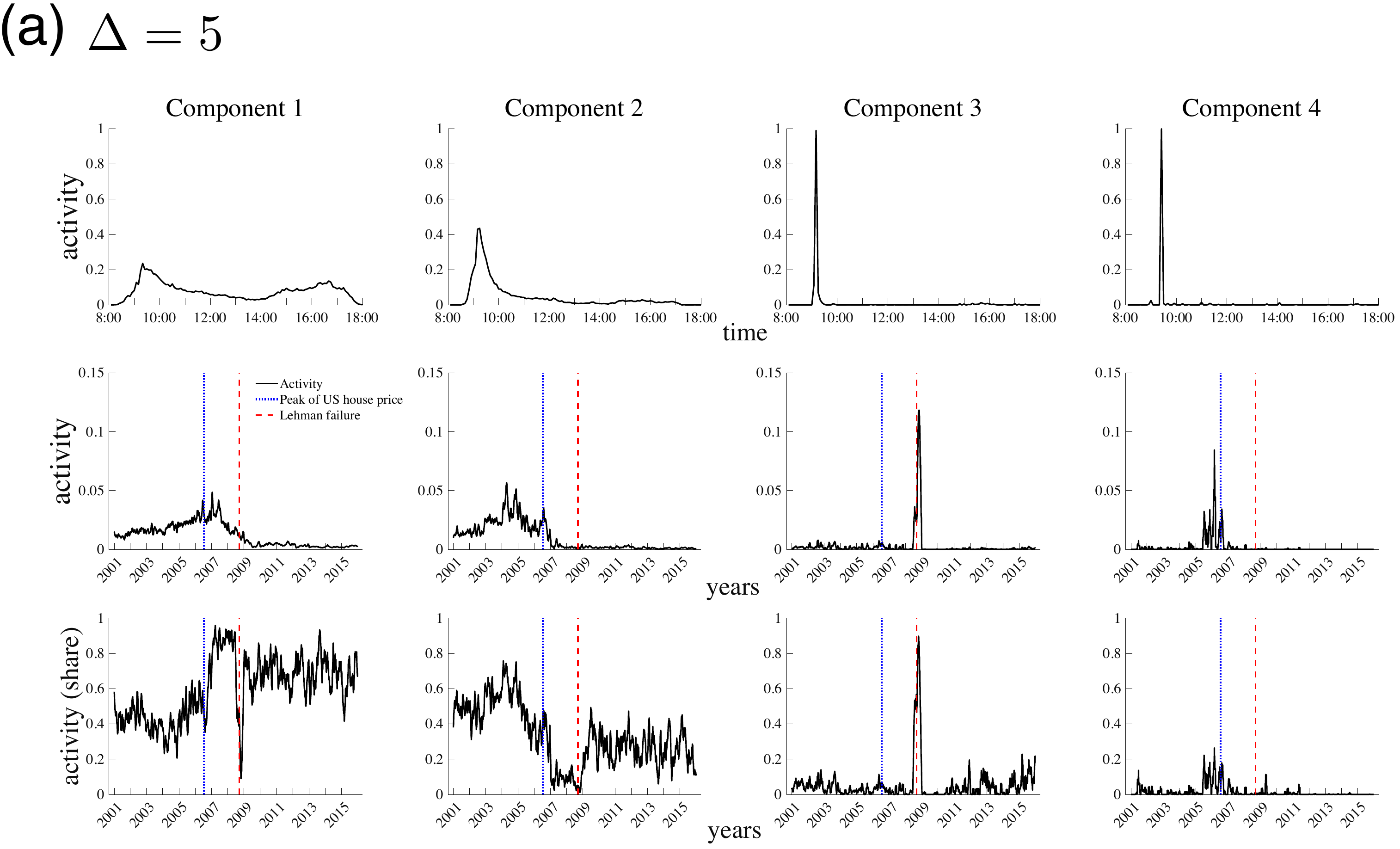}
     \caption{\footnotesize{Intra- and inter-day activities. Upper: Intra-day activity. Middle: Inter-day activity. Lower: Share of a component in the daily activity. (a) $\Delta = 5$, (b) $\Delta = 30$, and (c) $\Delta = 45$. See the caption of Fig.~\ref{fig:activity_resol15}.}}
     \label{fig:SI_activity} 
 \end{figure}

The $r$-th rows of factor matrices $\mathbf{B}$ and $\mathbf{C}$ (i.e., $\mathbf{b}_r$ and $\mathbf{c}_r$) respectively represent the intra- and inter-day activities of component $r$. Fig.~\ref{fig:SI_activity} illustrates intra- and inter-day activities for a given temporal resolution $\Delta\in\{5,30,45\}$. The selected rank size, based on the Core-Consistency measure, is $R = 4$ for $\Delta=5$, $R = 3$ for $\Delta=30$, and $R = 2$ for $\Delta=45$. We note that the choice of $\Delta$ does not affect the activity distributions for a fixed value of $R$. 
\bigskip
 
 \begin{figure}[ht]
     \centering
     \includegraphics[width=.8\columnwidth]{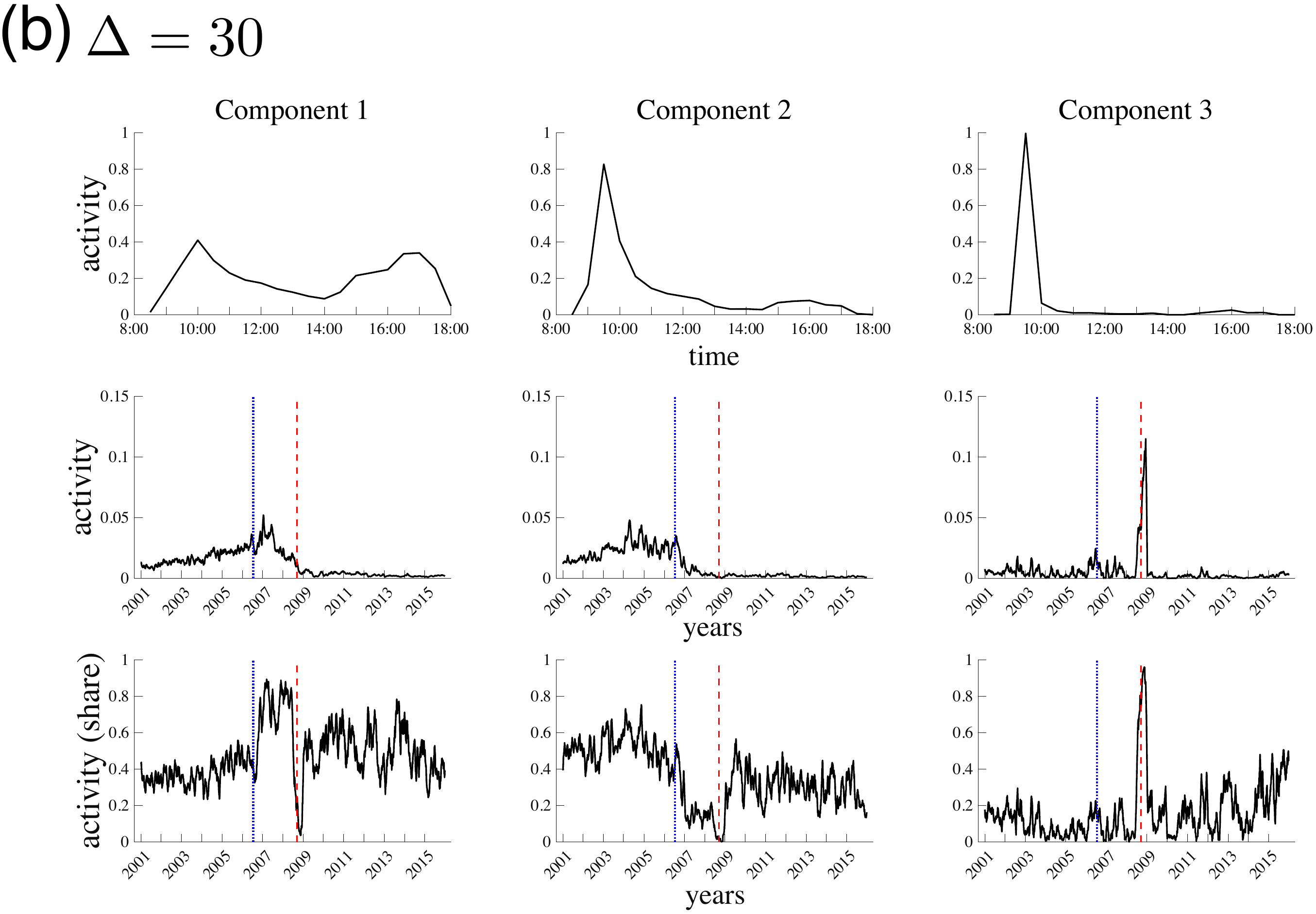}
 \end{figure}
 
 \begin{figure}[ht]
     \centering
     \includegraphics[width=.6\columnwidth]{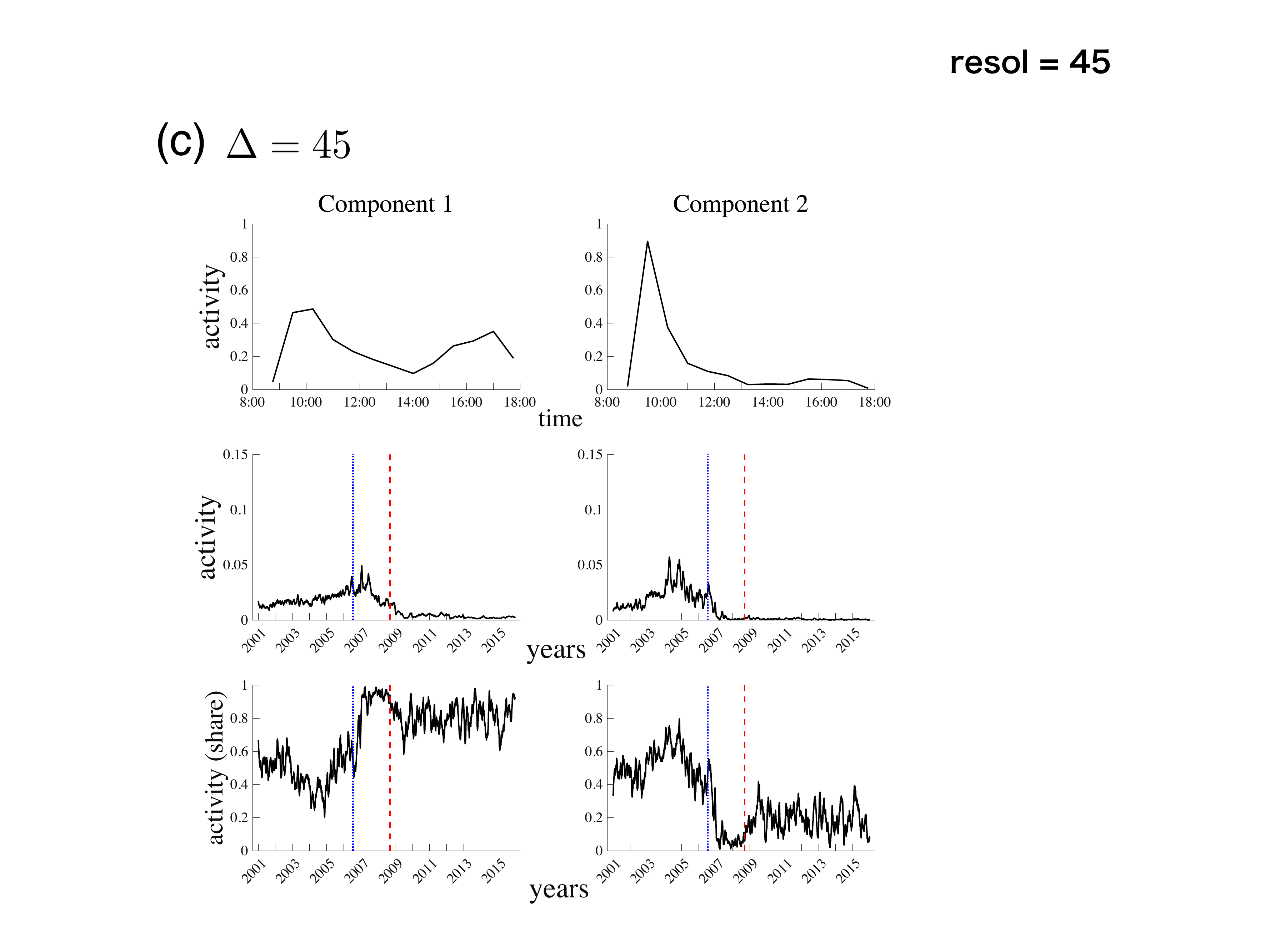}
 \end{figure}

\end{document}